\documentclass[a4paper,notitlepage,superscriptaddress,aps,preprintnumbers,amsmath,amssymb,sort&compress,nofootinbib,
]{revtex4-1}

\usepackage{amsfonts,amsmath,amssymb,bm,tensor}
\usepackage{comment}
\usepackage{ifpdf}
\usepackage{slashed}
\usepackage{color}
\usepackage[mathscr]{eucal}
\usepackage[utopia]{mathdesign}
\usepackage[utf8]{inputenc}

\usepackage{cancel}

\ifpdf
  \usepackage{graphicx}     
  \usepackage[bookmarksopen,colorlinks=true,linkcolor=bblue,citecolor=bblue,urlcolor=ppink]{hyperref}
\else     
  \usepackage[dvipdfmx]{graphicx}     
  \usepackage[dvipdfmx,bookmarksopen,colorlinks=true,linkcolor=bblue,citecolor=ppink,urlcolor=darkred]{hyperref}
\fi

\usepackage{natbib}

\definecolor{darkred}{rgb}{0.6,0,0}
\definecolor{darkgreen}{rgb}{0.992447,0.623778,0.034597}
\definecolor{ppink}{rgb}{1,0.4,0.4}
\definecolor{bblue}{rgb}{0.284602,0.317763,0.963947}
\definecolor{mygreen}{rgb}{0,0.7,0}
\definecolor{myred}{rgb}{1,0.3,0.4}
\definecolor{myblue}{rgb}{0.2,0.3,1}
\definecolor{refkey}{rgb}{0.3,0.6,1}
\definecolor{labelkey}{rgb}{1,0.1,1}

\usepackage{slashed}
\usepackage{braket}
\usepackage{comment}

\newcommand{\bs}{\boldsymbol}
\newcommand{\tx}{\text}
\newcommand{\mc}{\mathcal}
\newcommand{\al}[1]{\begin{align}#1\end{align}}

\newcommand{\ie}[1]{\begin{itemize}#1\end{itemize}}	
\newcommand{\ab}[1]{\left|#1\right|}

\newcommand{\prn}[1]{\left(#1\right)}
\newcommand{\sqbr}[1]{\left[#1\right]}

\newcommand{\bp}{\begin{pmatrix}}
\newcommand{\ep}{\end{pmatrix}}
\newcommand{\bb}{\begin{bmatrix}}
\newcommand{\eb}{\end{bmatrix}}

\newcommand{\pmat}[1]{\begin{pmatrix}#1\end{pmatrix}}

\newcommand{\Eq}[1]{Eq.(\ref{#1})}  
\newcommand{\Fig}[1]{Fig.\ref{#1}}

\newcommand{\appe}[1]{Appendix \ref{#1}}
\newcommand{\fn}[1]{\!\scalebox{0.8}{$\prn{#1}$} } 
\newcommand{\si}[1]{{ \!\scalebox{0.5}{$#1$} }} 

\newcommand{\nn}{\nonumber\\}
\newcommand{\qcq}{\quad,\quad}
\newcommand{\qcn}{\quad,\nn}

\newcommand{\ov}{\over}

\newcommand{\so}{\ensuremath{\Rightarrow}}

\DeclareMathOperator{\re}{Re}

\newcommand{\df}{\tx{d}}

\newcommand{\p}{\partial}

\def\sla#1{\rlap{\kern .15em /}#1}

\newcommand{\pr}{\prime}

\newcommand{\mpl}{M_{pl}}
\newcommand{\msol}{M_{\odot}}

\newcommand{\nm}{\mathrm{nm}}

\newcommand{\Mpc}{\mathrm{Mpc}}

\newcommand{\g}{\mathrm{g}}
\newcommand{\kg}{\mathrm{kg}}

\newcommand{\MeV}{\mathrm{MeV}}
\newcommand{\GeV}{\mathrm{GeV}}

\newcommand{\cs}[1]{\begin{cases}#1\end{cases}}
\definecolor{myGray}{rgb}{0.5,0.5,0.5}
\newcommand{\gray}[1]{{\color{myGray}#1}}
\newcommand{\Sec}[1]{Sec. \ref{#1}} 

\makeatletter
\newcommand\footnoteref[1]{\protected@xdef\@thefnmark{\ref{#1}}\@footnotemark}
\makeatother

\begin{document}

\title{Formation of primordial black holes in an axion-like curvaton model}

\author{Kenta Ando}
\affiliation{ICRR, University of Tokyo, Kashiwa, 277-8582, Japan}
\affiliation{Kavli IPMU (WPI), UTIAS, University of Tokyo, Kashiwa, 277-8583, Japan}
\author{Masahiro Kawasaki}
\affiliation{ICRR, University of Tokyo, Kashiwa, 277-8582, Japan}
\affiliation{Kavli IPMU (WPI), UTIAS, University of Tokyo, Kashiwa, 277-8583, Japan}
\author{Hiromasa Nakatsuka}
\affiliation{ICRR, University of Tokyo, Kashiwa, 277-8582, Japan}
\affiliation{Kavli IPMU (WPI), UTIAS, University of Tokyo, Kashiwa, 277-8583, Japan}

\begin{abstract}
\noindent

We perform the detailed analysis of the primordial black hole (PBH) formation mechanism in an axion-like curvaton model with a coupling to inflaton.
The phase direction of the complex scalar works as a curvaton and produces  enough PBHs to explain the black hole binaries ($\sim 30 \msol$) observed in the LIGO-Virgo Collaboration or PBHs as dark matter (DM) ($\sim 10^{-12} \msol$).
We examine whether our model satisfies the current constraints on the PBH mass spectrum, the curvature perturbation and the secondarily produced gravitational waves.
We also take into account ambiguity about the choice of the window functions and effect of the non-Gaussianity.

\end{abstract}
\preprint{IPMU18-0087}

\date{\today}
\maketitle
\tableofcontents



\section{Introduction}\label{sec: intro}

Since the first gravitational wave (GW) event was discovered in 2015, several binary black hole merger events have been observed by the LIGO-Virgo Collaboration~\cite{Abbott:2016blz,TheLIGOScientific:2016pea,Abbott:2016nmj,Abbott:2017vtc,Abbott:2017gyy,Abbott:2017oio}.
Many of these black holes (BHs) have masses around $30\msol(=60\times 10^{30}\kg)$.
It is in dispute how such heavy black hole binaries are formed by stellar evolution, and some authors suggest BH formation from the first Population III stars~\cite{Kinugawa:2014zha}.
Primordial black holes (PBHs) are another promising candidate for the origin of the binary BH mergers
~\cite{Bird:2016dcv,Clesse:2016vqa,Sasaki:2016jop,Eroshenko:2016hmn,Carr:2016drx}.
PBHs are produced by the gravitational collapse of overdense regions
in the radiation-dominated era or the early matter-dominated era~\cite{Hawking:1971ei,Carr:1974nx,Carr:1975qj}.

PBHs are also interesting as a dark matter (DM) candidate.
The microlensing observations give stringent constraints on the DM PBH mass and in particular, the observation with the Subaru Hyper Supreme-Cam (HSC) almost closed the DM mass window~\cite{Allsman:2000kg,Tisserand:2006zx,Wyrzykowski:2011tr,Niikura:2017zjd}. 
However, recently it has been found that the so-called ''wave effect'' 
\footnote{The observational wavelength in the Subaru/HSC($\sim 600\nm$,r -band) is smaller than the Schwarzschild radius of the lensing objects, PBHs lighter than $\sim10^{-10}\msol$.}
weakens the lensing effect and PBHs with mass $10^{-13}$--$10^{-10}\msol$ can still account for all DM of the Universe.

Since extensive research puts the severe constraints on the mass distribution of PBHs (see Figs.~\ref{powers},\ref{PBHsDistri}, and \ref{2ndGWfig} ), a PBH production mechanism should give a peak-like mass distribution to explain the LIGO events or DM successfully.
PBH production also requires the large amplitude of density perturbations in the small scales, $10^{-5}\,\Mpc$ for LIGO PBHs and $10^{-12}\,\Mpc$ for DM PBHs. 
Since observations of the cosmic microwave background (CMB) and large scale structure (LSS) show that the spectrum of the curvature perturbations on large scales $ 1\,\Mpc \sim 10^3 \,\Mpc$ is nearly scale invariant and its amplitude is small~\cite{Ade:2015xua,Nicholson:2009pi,Nicholson:2009zj,Bird:2010mp}, the large perturbations on small scales responsible for PBHs are hardly produced in simple single-field inflation models and may require a different mechanism, for example, multi-fields inflation \cite{Bugaev:2011wy}, double inflation \cite{Inomata:2017vxo}, curvaton dynamics \cite{Kohri:2012yw,Ando:2017veq} and other mechanisms \cite{Pi:2017gih,Garcia-Bellido:2016dkw}. 
	
In this paper we investigate an axion-like curvaton model~\cite{Kawasaki:2012wr,Kawasaki:2013xsa,Ando:2017veq} with an inflaton-coupling term. 
In the previous model~\cite{Ando:2017veq}, the complex scalar field has a large field value initially and the field value (more precisely the value of the radial direction) decreases during inflation.
Since the amplitude of the fluctuations of 
the curvaton ($=$ the phase direction of $\Phi$) is proportional to $1/|\Phi|$, a blue-tilted spectrum is produced.
Here we introduce a coupling between $\Phi$ and the inflaton which stabilizes the $\Phi$ at $\Phi\simeq 0$ for a large inflaton value.
When the scalar $\Phi$ is stabilized, the fluctuations of the curvaton decrease, which gives a blue-tilted spectrum on large scales. 
After the inflaton field value becomes some critical value, the stabilization due to the inflaton does not work and $\Phi$ rolls down to the true minimum. 
Thus, contrary to the model in ~\cite{Ando:2017veq}, the field value of $\Phi$ increases during inflation, which realizes a red-tilted spectrum of the curvaton fluctuations on smaller scales.
We consider the PBH production scenarios in two PBH mass cases, (1) DM PBHs with $10^{-13}\msol\sim 10^{-11}\msol$ and (2)LIGO PBH with $30 \msol$, and examine how the axion-like curvaton model can explain DM PBHs or LIGO PBHs without conflicting with current observational constraints.
In estimating the PBH abundance we take into account the effect of non-Gaussianity of the curvaton perturbations and ambiguity about the choice of the window functions.

In \Sec{sec:curvaton_model}, we explain the axion-like curvaton model with an inflaton-coupling term and derive the formula of the perturbation spectrum analytically.
In \Sec{sec:pbh_formation}, we review formulas about PBH formation including non-Gaussianity.
In \Sec{sec_2ndGW}, we summarize the evaluation of the secondary GWs in our calculation. 
In \Sec{sec_result}, we show the results of the numerical calculation for the power spectrum of the density perturbation, the secondary GWs, and the mass spectrum of PBHs.
\Sec{sec_conclu} is our conclusion.

\section{Curvaton model}
\label{sec:curvaton_model}

Here we describe our curvaton model. 
We introduce a complex scalar field $\Phi$ whose phase direction plays a role of the curvaton. 
The complex scalar $\Phi$ has a Higgs-like potential, a coupling with an inflaton field $\phi$ and a linear term as
\al{
V_{\Phi}=\frac{\lambda}{4} 
	\prn{\ab{\Phi}^2
	-{v^2\ov 2}
  		  }^2 
    +g \phi^2 \ab{\Phi}^2
	- v^3\epsilon(\Phi+\Phi^*),
\label{potecomp}
}
where $\lambda$ and $g$ are coupling constants, $v/\sqrt{2}$ is the vacuum expectation value after inflation, and we assume that the last term is small ($\epsilon \ll 1$).\footnote{
{The small $\epsilon$ is natural in the sense of 't Hooft's naturalness~\cite{tHooft:1979rat} since $U(1)$ symmetry is restored for $\epsilon =0$.}}

We assume that the inflaton field value decreases during inflation.
Then in the early stage of inflation ($\phi \gtrsim (\lambda/g)^{1/2}v$), the coupling term with the inflaton fixes $\Phi$ near the origin.
In the late stage, the inflaton field value $\phi$ becomes smaller and $\Phi$ starts to roll down the Higgs-like potential. 
If $\Phi$ rolls down from the {\it exact} origin, quantum fluctuation of $\Phi$ given by $H/(2\pi)$ ( $H$: Hubble parameter) affects its dynamics and hence the evolution of $\Phi$ does not follow the classical equation of motion.
In such a case we should use the stochastic formalism for the complex scalar dynamics~\cite{Kawasaki:2015ppx}. 
To avoid this complexity, we add the small linear term $- v^3\epsilon(\Phi+\Phi^*) $ to the potential. 
The linear term slightly shifts the stabilized field value of $\Phi$ as $\Phi \sim \epsilon v/\lambda$ which is assumed to be a few Hubble away from the origin.
Then the scalar $\Phi$ rolls down the potential quickly and the classical dynamics dominate the quantum fluctuations. 
Moreover, the small shift term also avoids the cosmic string problem.  

Suppose that the scalar $\Phi$ rolls down the Higgs-like potential at $t_\text{pbh}$.
Here the subscript {''pbh''} means the horizon crossing of the perturbation with the PBH scale $k_\tx{pbh}\simeq 10^5 \,\Mpc^{-1}$ (LIGO-PBH) and $10^{12} \,\Mpc^{-1}$ (DM-PBH) during inflation. 
The effective mass of $\Phi$ is given by
\al{
    m_{\varphi}^2\sqbr{\ab{\Phi},\phi}
    =\frac{1}{2}\frac{\p^2}{\p \ab{\Phi}^2}V_{\Phi}
    =\frac{\lambda}{2}
    	\prn{3\ab{\Phi}^2-{v^2\ov 2}}+g\phi^2.	
	\label{mvarphi}
}
$\Phi$ starts to roll down when $m_{\varphi}\sqbr{\ab{\Phi}=0,\phi=\phi\fn{t_\tx{pbh}}}\simeq0$ or 
\al{
g\simeq
	\frac{\lambda}{4}
	\prn{\frac{v}{\phi\fn{t_\tx{pbh}}}}^2.
\label{gasum}
}
In our calculation, we assume that the effective mass is much larger than the Hubble parameter during inflation $\lambda v^2\gg H^2$ to achieve the large tilt of the power spectrum.
Since the curvaton is trapped by the steep potential until $t_\text{pbh}$, the choice of the initial condition hardly affects the results.

Let us evaluate the fluctuations of the complex scalar $\Phi$ which leads to the curvature perturbations.  
A peak-like spectrum of the perturbations requires a red-tilted shape at $k_\tx{pbh}<k$ and a blue-tilted shape at $k<k_\tx{pbh}$. 
We explain the mechanism of generating tilts in the following.
We temporarily write $\Phi$ as 
$\Phi=1/\sqrt{2}(\varphi_0+\varphi) e^{i\theta }$
with the homogeneous solution $\varphi_0$, and the perturbations $\varphi$ and $\theta$.
We define the canonical field 
$\tilde \sigma=\varphi_0\theta$.

During inflation, $\varphi$ and $\tilde \sigma$ acquire fluctuations with amplitude $H/(2\pi)$ at horizon crossing $t_k$,
\al{
&\prn{\frac{H}{2\pi}}^2
	=\mc P_{\varphi}\fn{k,t_k}
    =\mathcal{P}_{\tilde\sigma}(t, t_k)
	={\varphi_0^2\fn{t_k}}\mc P_{\theta}\fn{k,t_k}
\quad\so\quad
\mc P_{\theta}\fn{k,t_k}
	= \prn{\frac{H\fn{t_k}}{2\pi \varphi_0\fn{t_k}}}^2 .
\label{anguHubble}
}
After the horizon crossing, the evolution of the fluctuations depends on the effective mass.
The field $\tilde \sigma$ obtains the effective mass $\tilde{m}_\sigma$ through the linear term as
\al{
-v^3 \epsilon \prn{\Phi+\Phi^*}
	&=-2{v^3\epsilon}   
    {\varphi_0 \ov\sqrt{2}}
    \cos \left(\frac{\tilde \sigma}{{\varphi_0}}\right)
	\simeq 
    	-2{v^3\epsilon} 
        	{\varphi_0 \ov\sqrt{2}}
    	+{v^3\epsilon} 
       {1 \ov\sqrt{2}\varphi_0}
        \tilde \sigma^2
    = -2v^3 \epsilon\Phi_0 
    	+{1\ov 2}\tilde{m}_\sigma \tilde \sigma^2,
  \qcn
\tilde{m}_\sigma^2&= 
 {\sqrt{2} \epsilon v^2\frac{v}{\varphi_0} } .
 \label{msigma}
} 
When the scalar field is stabilized near the origin, $\varphi_0 \sim \epsilon v/\lambda$. 
We choose the model parameters to achieve $\varphi_0 \gtrsim H$ and $\tilde m_\sigma^2> H^2$ at $t < t_\text{pbh}$ and $\tilde m_\sigma^2<H^2$ for $\varphi_0 \sim v$.
Using the equation of motion of the perturbation in \Eq{damphigh}, we get the damping factor on $\tilde \sigma$ as
\al{
\ln R_k&
	=\ln 
    \prn{\frac{\tilde\sigma_k\fn{t_\tx{end}}}{\tilde\sigma_k\fn{t_k}}}
	=-\frac{3}{2}\int^{t_\text{pbh}}_{t_k} H
   \re \sqbr{1-\sqrt{1-\prn{\frac{2\tilde m_\sigma}{3H} }^2}}\df t
	,
}
where $t_\text{end}$ denotes the time when inflation ends.
Since $\tilde m_\sigma$ is larger than Hubble before {$\varphi_0$} rolls down, the damping factor for $k< k_\text{pbh}$ is approximately written as 
\al{
\ln R_k\simeq 
	-\frac{3}{2}\int^{t_\text{pbh}}_{t_k} H\df t
	=-\frac{3}{2}\prn{N\fn{t_\text{pbh}}-N\fn{t_{k}}}
	\sim
	\ln\prn{\frac{k}{k_\text{pbh}}}^{3/2}
\label{Dampsim}
}
where $N$ is $e$-foldings defined by $N\fn{t}=\ln\prn{{a(\!t_\tx{end}\!)}/{a(\!t\!)}}$.
Thus, for the perturbation of $\tilde\sigma$ with $k< k_\text{pbh}$, the power spectrum is blue-tilted $\mc P_{\tilde\sigma}$ (and hence $\mathcal{P}_\theta$) $\propto k^{3}$.
On the other hand, since ${\varphi_0\fn{t} }$ grows exponentially with time for $t \gtrsim t_\text{pbh}$ (see \Fig{Zeromodedynamisc}), $\mc P_{\theta}\fn{k,t_k}$ is sharply red-tilted at  $k> k_\tx{pbh}$.
As for the radial direction, $\varphi$ mostly has large positive mass $m_{\varphi}[\ab{\Phi},\phi]$ compared to Hubble parameter during inflation. Thus, $\varphi$ is highly suppressed and negligibly small.
After inflation, only the angular perturbation is left. 

We assume that {$\tilde \sigma$} obtains the axion-like potential through some nonperturbative effect as
\al{
V_\sigma
	=\Lambda^4
    \sqbr{1-\cos\prn{
    	{\tilde \sigma\ov v}
        -\theta_i
        }	 }
   \simeq {1\ov 2}m_\sigma^2 
   (\tilde \sigma-v\theta_i)^2
    ={1\ov 2}m_\sigma^2 v^2(\theta-\theta_i)^2
    \label{eq:axion_pot}
} 	
with $m_\sigma=\Lambda^2/v$ and the misalignment angle $\theta_i$.\footnote{
The minimum of phase direction of the potential is determined by the linear term for ${\varphi}_0 \ll v$. 
This minimum, in general, is different from the minimum of Eq.~(\ref{eq:axion_pot}), which results in the misalignment angle $\theta_i$. 
} 
After $\varphi_0\to v$, we define the curvaton as 
\al{
\sigma=\tilde \sigma -v \theta_i 
	=v\delta\!\theta.
}
with $\delta\!\theta=\theta-\theta_i$.
The density perturbation of the curvaton is given by
\al{
{\delta \rho_{\sigma}\ov\rho_\sigma}=2{\sigma\ov v\theta_i}=2{\delta \!\theta\ov \theta_i}.
\label{densityperturb}
}

Now let us calculate the curvature perturbations $\zeta$.
The density perturbation contains both contributions from the inflaton and the curvaton . 
With the energy ratio of the curvaton to the inflaton $r=\rho_\sigma/\rho_I$,  the curvature perturbation is given by
\al{
\zeta
&= -{H\ov \dot\rho }\delta \rho
=-{H\ov \dot\rho_I+\dot\rho_\sigma }(\delta \rho_I+\delta \rho_\sigma)
={\dot\rho_I\ov \dot\rho_I+\dot\rho_\sigma }\prn{-H\delta\rho_I\ov \dot\rho_I}
	+{\dot\rho_\sigma\ov \dot\rho_I+\dot\rho_\sigma }\prn{-H\delta\rho_\sigma\ov\dot\rho_\sigma}
\nn&=
	{4\ov 4+3r }\prn{-H\delta\rho_I\ov \dot\rho_I}
	+{3r\ov4+3r}\prn{-H\delta\rho_\sigma\ov\dot\rho_\sigma}
\label{zetadens}
}
where we use $\dot\rho_I=-4H\rho_I$ and $\dot\rho_\sigma=-3H\rho_\sigma$.
From Eqs.~(\ref{Dampsim})-(\ref{zetadens}),  neglecting the contribution from the inflaton, the curvature perturbation is finally given by 
\al{
\mc P_{\zeta}\fn{k,t_\tx{end}}
	&=
	\prn{3r\ov4+3r}^2
	\prn{2\ov 3\theta_i}^2
	\mc P_{\theta}\fn{k,t_k}
	R_k^2
\nn	&=
	\prn{2r\ov4+3r}^2
	\prn{H\fn{t_k}\ov 2\pi {\varphi_0\fn{t_k}}\theta_i}^2
	R_k^2.
	\label{Pzetaformula}
}
We are interested in the curvature perturbation after the curvaton decay into the radiation.
Although $r$ is small during inflation, $r$ grows after inflation since the curvaton behaves as matter-like
$\rho_\sigma\propto a^{-3}$ until the curvaton decay.
 In the following, $r$ denotes the ratio at the curvaton decay. 
 We make a list of typical parametrization in Table \ref{rthetaTable} and \Eq{curvatonParames}.
 With $r\sim \mc O\fn{0.1}$, we take the curvaton decay temperature $T_\sigma\sim 10^7\,\GeV$
 (see the detailed discussion in \appe{curvatondynamics}).

\section{PBH production}
\label{sec:pbh_formation}

In our model, the PBH production occurs during the radiation-dominated era. We briefly summarize the useful formulas to calculate PBH mass distribution. 
In the radiation-dominated era, overdensity regions collapse into black holes when the scale of density fluctuations re-enters the horizon.
Thus, the PBH mass is roughly given by the horizon mass at that time and is related to the scale of the perturbation $k$ or the formation temperature $T$ as~\cite{Inomata:2017vxo}
 \al{
 M\fn{k}
&=
	\gamma \rho_r \frac{4\pi}{3}H^{-3} \bigg|_{k=aH},
\nn&\simeq
	10^{-12}\msol \prn{\gamma\ov 0.2}\prn{g_*\ov 106.75}^{-1/6}
	\prn{k\ov 1.55\times 10^{12}\,\Mpc^{-1}}^{-2},
\nn&\simeq
	10^{-12} \msol \prn{\gamma\ov 0.2}\prn{g_*\ov 106.75}^{-1/2}
	\prn{T\ov  9.75\times 10^4\,\GeV }^{-2},
\nn&\simeq
	30 \msol \prn{\gamma\ov 0.2}\prn{g_*\ov 10.75}^{-1/6}
	\prn{k\ov 3.43\times 10^{5}\,\Mpc^{-1}}^{-2},
\nn&\simeq
	30 \msol \prn{\gamma\ov 0.2}\prn{g_*\ov 10.75}^{-1/2}
	\prn{T\ov 31.6\,\MeV }^{-2},
	\label{PBHmasses}
}
where $\rho_r$ is the radiation energy density and $\gamma$ is the ratio of PBH mass to the horizon mass. 
We use the simple analytical estimation  $\gamma=3^{-3/2}\simeq 0.2$ \cite{Carr:1975qj} in this paper. 
  
The power spectrum of the curvature perturbations $\mc P_\zeta$ determines the PBH production rate. 
At first, we assume the curvature perturbations follow the Gaussian statistics. 
We will take non-Gaussianity into account later.
PBH production depends on the coarse-grained density perturbation over the horizon. 
The density perturbation in comoving gauge is related to the curvature perturbation as 
$\delta=(4/9)(k/aH)^2\zeta$.
Once the coarse-grained density perturbation exceeds the threshold value $\delta_c$, the horizon mass collapses into black holes.
In this paper, we take the threshold as $\delta_c=0.4$ \cite{Harada:2013epa}. 
The coarse-grained density perturbation is given by
\al{
\delta_W\fn{\bs x;R}
	&=
	\int \df ^3 y W\fn{\ab{\bs x-\bs y};R}\delta (\bs y)
\nn&	=
	\int {\df ^3 k\ov (2\pi)^3} \tilde W\fn{k;R} e^{i\bs k\cdot \bs x} \delta_k
}
and its correlation function is
\al{
\braket{\delta_W^2}\fn{M\fn{k}}
	=\braket{\delta_W\fn{\bs x;k^{-1}} \delta_W\fn{\bs x;k^{-1}}	}
	&=\int \df \ln q \tilde W^2\fn{q;k^{-1}}{16\ov 81}\prn{q/k}^4 \mc P_\zeta\fn{q}T\fn{q,k^{-1}}^2
} 
where $\tilde W\fn{k;R}$ is a window function in momentum space and $T\fn{k,\eta}$ is the transfer function. 
Since we assume that the curvaton has already decayed into the radiation before the PBH production, we use the transfer function in the radiation-dominated era,
\al{
T\fn{k,\eta}
	=3{\sin\fn{k\eta/ \sqrt{3}}-{k\eta/ \sqrt{3}}\cos\fn{k\eta/ \sqrt{3}}\ov \prn{k\eta/ \sqrt{3}}^3}.
}
Note that the choice of the window function causes $\mc O\fn{1}$ uncertainties on $\braket{\delta_W^2}$~\cite{Ando:2018qdb}.
We calculate $\braket{\delta_W^2}$ using Gaussian type $\tilde W\fn{k,R}=e^{-(kR)^2/2}$, real-space top-hat type $\tilde W\fn{k;R}= 3\prn{\sin(kR)-kR\cos(kR)}/(kR)^3$ and 
delta-function type $\tilde W^2\fn{k;R}=\delta \fn{kR-1}$. 
For the scale invariant case $\mc P_{\zeta}\fn{k}=A_s$, $\braket{\delta_W^2}$ depends on the choice of the window function as~\cite{Ando:2018qdb}
\al{
\braket{\delta_W^2}=
\cs{
	1.06~A_s \quad&\tx{(real-space top-hat)}\\
	0.191~A_s\quad&\tx{(delta-function)}\\
	0.0867~A_s \quad&\tx{(Gaussian)}.
	}
	\label{windowChoie}
}
Once $\braket{\delta_W^2}$ is given, one can estimate the PBH production rate as
\al{
\beta\fn{M}
	=\int_{\delta_c}^\infty
	{\df \delta\ov\sqrt{2\pi \braket{\delta_W^2\fn{M}}}}
	e^{-{\delta^2\ov2\braket{\delta^2_\si{~W}\fn{M}}}}
	&\simeq {1\ov \sqrt{2\pi}}{\sqrt{\braket{\delta^2_{W}(M)}}\ov\delta_c }e^{-{\delta_c^2\ov 2\braket{\delta^2_\si{~W}\fn{M}}}}.
} 
,which approximately follows log-normal distribution.
Our calculation of $\beta\fn{M}$ is based on the conventional Press-Schechter formalism.
Recently, detailed calculations were discussed based on the peak theory
\cite{Germani:2018jgr,Yoo:2018kvb}. 
 
The mass spectrum of PBH is given by~\cite{Inomata:2017vxo}
\al{
f\fn{M}
	&={\df\Omega_\tx{PBH}\ov\df \ln M }{1\ov \Omega_\tx{DM} }
	={\gamma\rho_r\fn{t_\tx{M}}\beta\fn{M}\ov \rho_\tx{m}\fn{t_\tx{M}} }{\Omega_\tx{m}\ov \Omega_\tx{DM} }
\nn	&\simeq
    \prn{\beta\fn{M}\ov 1.0\times 10^{-14}}
	\prn{\gamma\ov 0.2}^{3/2}
	\prn{106.75\ov g_*\fn{T_\tx{M}}}^{1/4}
	\prn{0.12\ov \Omega_\tx{DM}h^2}
	\prn{M\ov 10^{-13}\msol}^{-1/2}
\nn	&\simeq
    \prn{\beta\fn{M}\ov 1.8 \times 10^{-8}}
	\prn{\gamma\ov 0.2}^{3/2}
	\prn{10.75\ov g_*\fn{T_\tx{M}}}^{1/4}
	\prn{0.12\ov \Omega_\tx{DM}h^2}
	\prn{M\ov \msol}^{-1/2}
	.
	\label{fMformula}
}
Here the subscription "m" denotes the matter (baryon+DM).
With use of $f\fn{M}$, the total fraction of DM in PBH is rewritten as
\al{
{\Omega_\tx{PBH}\ov \Omega_\tx{DM} }=\int \df \ln M f\fn{M}.
}
It is known that ${\Omega_\tx{PBH}/\Omega_\tx{DM} }\simeq 10^{-3}$ explains the event rate of the binary black hole mergers observed by the LIGO~\cite{Sasaki:2016jop}.
Note that the above-mentioned PBH production mechanism has some uncertainties on $\gamma$, $\delta_c$ and the choice of window functions. 
In this paper, we choose the conservative values for $\gamma$ and $\delta_c$.

So far we have assumed that the curvature perturbations are Gaussian. 
Now we take into account the effect of non-Gaussianity.
In fact, it is known that the curvaton produces significant non-Gaussianity in curvature perturbations.
Here we briefly summarize our treatment of the non-Gaussianity in this paper based on \cite{Byrnes:2012yx,Young:2013oia}.

Non-Gaussian distribution with the local type bispectrum can be written as
\al{
\zeta\fn{x}=\zeta_g\fn{x} +{3\ov 5}f_\text{NL}\prn{\zeta_g^2\fn{x}-\braket{\zeta^2_g\fn{x}}}
\label{zetadeco}
}
where $\zeta_g\fn{x}$ follows a Gaussian distribution. In curvaton models,  the non-Gaussianity parameter $f_\text{NL}$ is determined by the ratio $r$ of 
the curvaton density to radiation density at the curvaton decay and is written as
\al{
f_{NL}={5\ov 12}\prn{-3+{4\ov r}+{8\ov 4+3r}}.
}
There is one difficulty in considering the non-Gaussian effect.
The $f_\text{NL}$ is defined in the curvature perturbation $\zeta$ while the PBH formation is calculated by the coarse-grained comoving density perturbation.
Here we recalculate $\beta(M)$ using $\zeta$ at the horizon crossing and estimate the non-Gaussian effect.

Non-Gaussianity modifies our discussion in two points.
First, non-Gaussianity could increase the PBH fraction $\beta\fn{M}$ by amplifying the probability at $\delta \simeq \delta_c$ of the distribution function.
Using~\Eq{zetadeco},
the PBH fraction including non-Gaussianity $\beta\fn{\braket{\zeta^2_g},f_{NL}}$ is 
estimated as~\cite{Young:2013oia}
\al{
\beta\fn{\braket{\zeta^2_g},f_{NL}}
	&
\simeq
	{1\ov \sqrt{2\pi}}
	\prn{
	{1\ov y_{c+} }e^{-{y_{c+}^2\ov 2}}
	+{1\ov y_{c-} }e^{-{y_{c-}^2\ov 2}}
	}
}
where the $y_{c\pm}$ is given by
\al{
y_{c\pm}={1\ov \sqrt{\braket{\zeta^2_g}}}
	{5\ov 6f_{NL}} 
	\sqbr{-1\pm 
		\sqrt{1+{12\ov 5}f_{NL}\prn{{3\ov 5}f_{NL}\braket{\zeta_g^2}+\zeta_c}}
		}.
}
Here the critical curvature $\zeta_c$ depends on the critical density $\delta_c=0.4$ and the choice of window functions.
We approximately use \Eq{windowChoie} as the relation between $\zeta_c$ and $\delta_c$:
\al{
\zeta_c =
\cs{
	0.389 \quad&\tx{(real-space top-hat)}\\
	0.915\quad&\tx{(delta-function)}\\
	1.36 \quad&\tx{(Gaussian)}.
	}
}
Since the curvature perturbation with $k$ at the peak of the power spectrum dominantly contributes to the PBH production, we approximate $\braket{\zeta_g^2}$ as the peak value $\mc P_{\zeta_g}\fn{k_\tx{pbh}}$.
For a given PBH fraction $\beta_c$, $f_{NL}$ effectively lowers the required curvature perturbation $\mc P_{\zeta_g}\fn{k_\tx{pbh}}\to B(f_{NL})\mc P_{\zeta_g}$ given by
\al{
\beta_c
	=\beta\fn{\mc P_{\zeta_g},f_{NL}=0}
	=\beta\fn{B(f_{NL})\mc P_{\zeta_g},f_{NL}}.
	\label{Deampnongau}
}

The second effect of non-Gaussianity is that the power spectrum obtains an additional contribution from the second term in Eq.~(\ref{zetadeco}), which is written as
\al{
\mc P_\zeta\fn{k}
	&=\mc P_{\zeta_g}\fn{k}
	+	\prn{{3\ov 5}f_{NL}}^2
	{k^3\ov 2\pi}\int{\df^3 q}{1\ov q^3} {1\ov \ab{\bs k-\bs q}^3}
	\mc P_{\zeta_g}\fn{\bs q}\mc P_{\zeta_g}\fn{\ab{\bs k-\bs q}}.
}
We define the amplification factor of the power spectrum at the peak value as
\al{
\sqrt{Q\fn{\mc P_{\zeta_g}\fn{k_\tx{pbh}},f_{NL}}}
	={\mc P_\zeta\fn{k_\tx{pbh}}\ov \mc P_{\zeta_g}\fn{k_\tx{pbh}}}.
\label{amplPzeta}
}
In the following calculation, we include the effect of non-Gaussianity on both the PBH production and perturbations.

\section{Constraints of the secondary gravitational wave }
\label{sec_2ndGW}

The large curvature perturbations produce the secondary GWs~\cite{Ananda:2006af,Baumann:2007zm,Saito:2008jc,Saito:2009jt,Bugaev:2009zh,Bugaev:2010bb}.
This fact should be taken into account when we discuss models of inflationary PBHs since such GWs are severely constrained by the Pulsar Timing Array (PTA) experiments as mentioned later (see Fig.~\ref{2ndGWfig})~\cite{Kawasaki:2013xsa,Inomata:2016rbd,Garcia-Bellido:2017aan,Orlofsky:2016vbd}.  
We calculate the differential density parameter of the produced GWs. 
The GWs are mainly produced when the peak wavelength of the curvature perturbation spectrum re-enters the horizon in the radiation-dominated era.
After production, the GW energy density decreases by the cosmic expansion, and at present the density parameter of the produced GWs is given by~\cite{Ando:2018qdb}
\al{
\Omega_{GW}\fn{k,t_{0}}
	=\prn{a_c^2 H_c\ov a_0^2 H_0}^2\Omega_{GW}\fn{\eta_c,k}
\simeq 
	0.83
		\prn{\tx{g}_c\ov 10.75}^{-1/3}
		\Omega_{r,0}
		\Omega_{GW}\fn{\eta_c,k},
\label{2ndGWformula}
}
where $\eta_c$ is some time when the secondary GW generation effectively finishes and  the GWs behave as radiation $\rho_{GW}\propto a^{-4}$ for $\eta >\eta_c$.
Here $g_c$ is the degrees of freedom of the radiation at $\eta_c$.
In our case, we assume $g_c=10.75$ for LIGO-PBHs and $g_c=106.75$ for DM-PBHs, respectively.
$\Omega_{GW}\fn{\eta_c,k}$ is calculated using the power spectrum of the curvature perturbations $\mc P_\zeta\fn{k}$ as~\cite{Inomata:2017vxo}
\al{
\Omega_{GW}\fn{\eta_c,k}
&=
\Braket{
	{8\ov 243}
	\int^\infty_0\df y
	\int^{1+y}_{\ab{1-y}}\df x
	{y^2\ov x^2}
	\prn{1-{(1+y^2-x^2)^2\ov 4y^2}}^2
	\mc P_\zeta\fn{kx}\mc P_\zeta\fn{ky}
	\sqbr{
		{k^2\ov a\fn{\eta_c}}
		\int^{\eta_c}\df \bar\eta
		a\fn{\bar\eta}
		g_k\fn{\eta_c;\bar\eta}
		f\fn{ky,kx,\bar\eta}
		}^2
        },
        \label{eq:gravitational_wave}
}
where  $\braket{..}$ means the time average over $\eta_c$
, $T\fn{\eta,k}$ is the transfer function of the radiation and $g_k$ is the Green function.
$T\fn{\eta,k}$ and $g_k$ are given by
\al{
T\fn{\eta,k}&
	=9\sqrt{3}{\sin(k\eta/\sqrt{3})-(k\eta/\sqrt{3})\cos (k\eta/\sqrt{3})\ov (k\eta)^3}
\qcn
g_k\fn{\eta,\tilde\eta}
	&={\sin\fn{k(\eta-\bar\eta)}\ov k}\theta \fn{\eta-\bar\eta}.
}
$f\fn{k_1,k_2,\eta}$ is written as
\al{
f\fn{ k_1,k_2,\eta}
	&=
	\sqbr{
	2T\fn{k_1,\eta}T\fn{k_2,\eta}
	+\prn{{\dot T\fn{k_1,\eta}\ov H\fn{\eta}}+T\fn{k_1,\eta}}
	\prn{{\dot T\fn{k_2,\eta}\ov H\fn{\eta}}+T\fn{k_2,\eta}}
	}.
}
Considering the non-Gaussianity, we multiply the factor 
$B(f_\text{NL})^2\,Q\fn{B(f_\text{NL})\mc P_{\zeta_g}\fn{k_\tx{pbh}},f_{NL}}$
[see, Eq.(\ref{Deampnongau}) and \Eq{amplPzeta}] to the $\Omega_{GW}$.

\section{Results}\label{sec_result}

\subsection{Classical dynamics of the complex scalar}

During inflation, $\Phi_0{=\varphi_0/\sqrt{2}}$ obeys the equation of motion~\Eq{Csol}. 
We numerically solve it and obtain the classical dynamics for parameters which are appropriate for PBH-DMs and LIGO-PBHs. 
As for the inflation model, for simplicity, we adopt the chaotic inflation with potential $V_\text{inf}=m_\phi^2\phi^2/2$ with $m_\phi=(5\times 10^{-6}\mpl )\simeq 10^{13}\,\GeV$ ($\mpl$:reduced Planck mass). 
We take the initial inflaton value $\phi_{in}= 15.6 \mpl$ which achieves the $e$-foldings $N>55$. 

We take the typical parameter for PBH-DM and LIGO-PBH as 
\al{
&	{v_\si{DM}=4.84\times 10^{-2}\mpl }
\qcq
	&&\lambda_\si{DM}=5\times 10^{-5}
\qcq
	&&& {\epsilon_\si{DM}=2.81\times 10^{-10} }
\qcq
	&&&&g_\si{DM}=4.2\times 10^{-10}
\nn&	{ v_\si{LIGO}\simeq 8.10\times 10^{-2}\mpl }
\qcq
	&&\lambda_\si{LIGO}=5\times 10^{-5}
\qcq
	&&&{\epsilon_\si{LIGO}=1.08\times 10^{-10} }
\qcq
	&&&&g_\si{LIGO}=7.08\times 10^{-10}
\label{curvatonParames}
}
and we slightly change the parameter $v_\si{LIGO}$ for each window selection on LIGO-PBH since the different window function produces PBHs with different masses.

\begin{figure}[ht]
\centering 
\includegraphics[width=.65\textwidth]{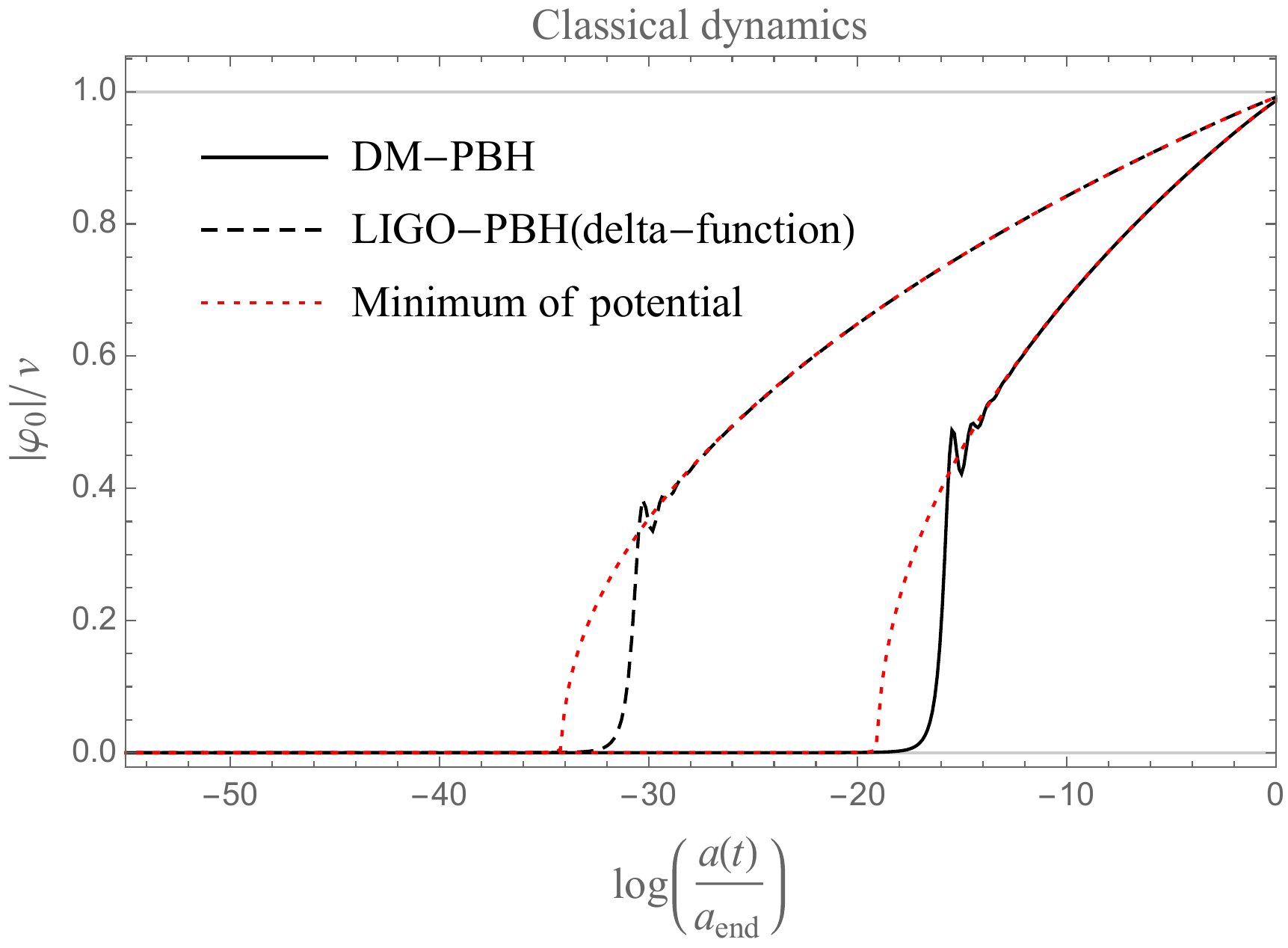}
\caption{
 	 Time evolution of $\ab{\Phi_0}$ during inflation for DM-PBH (black solid line) and LIGO-DM (black dashed line).
  	The field value is normalized with $v$. 
  	The horizontal axis denotes the scale factor normalized with the inflation end $\log (a(t)/a_\text{end})$.
	The red lines show the potential minimum for DM-PBH and LIGO-DM. 
}
 \label{Zeromodedynamisc}
\end{figure}

We show the results in \Fig{Zeromodedynamisc}.
In the early stage of inflation, $\Phi_0$ is fixed near the origin due to the inflaton-coupling $g\phi^2\ab{\Phi}^2$. 
After the perturbation with the PBH scale crosses the horizon, the 
effective mass near the origin becomes tachyonic and $\Phi_0$ rolls down the Higgs-like potential. 
Since we introduce $U(1)$ symmetry breaking linear term $- v^3\epsilon(\Phi+\Phi^*)$, the complex scalar field rolls down to $\arg\fn{\Phi}=0$ direction. 
The red lines in \Fig{Zeromodedynamisc} show the minimum of the potential.
At $t_\tx{pbh}$, the potential minimum shifts toward $v$. At that moment, $\Phi_0$ still stays near the origin.
Because of the small linear term, $\Phi_0$ has nonzero value and the field value grows quickly to follow the potential. 
 When $\Phi_0$ catches up with the potential minimum, $\Phi_0$ oscillates around the minimum.
In \Fig{Zeromodedynamisc}, the oscillating behavior around the potential minimum is seen at $\log(a/a_\text{end})\simeq -30$ for LIGO-PBH and $\simeq -16$ for DM-PBH. 
Finally, the inflation-coupling vanishes and the potential minimum becomes $v$.


We also check the consistency of the calculation.
In the previous calculation, we assumed that the quantum fluctuations do not disturb the classical dynamics, which is justified if
\al{
{\ab{\Phi_0}\ov   H/(2\pi)} >1
\qcq
{\ab{\dot \Phi_0}\ov H^2/(2\pi ) }>1
}
when $\Phi_0$ starts to roll down. We numerically confirm that $2\pi{\ab{\Phi_0}/ H}>3$ and $2\pi {\ab{\dot \Phi_0}/ H^2}>2$ at the time $m^2_\varphi=0$. 

In our numerical calculation, we include the energy density of the complex scalar \Eq{potecomp} into the total energy density of the Universe during inflation. 
During inflation, the complex scalar slightly affects the inflation dynamics.
The energy ratio of the complex scalar to the inflaton has the maximum value (17\%) at $t_\text{pbh}$ but decreases quickly as the complex scalar rolls down the Higgs-like potential. 


\subsection{The perturbation of the curvaton}

\begin{figure}[t]
 \centering 
  \includegraphics[width=.65\textwidth]{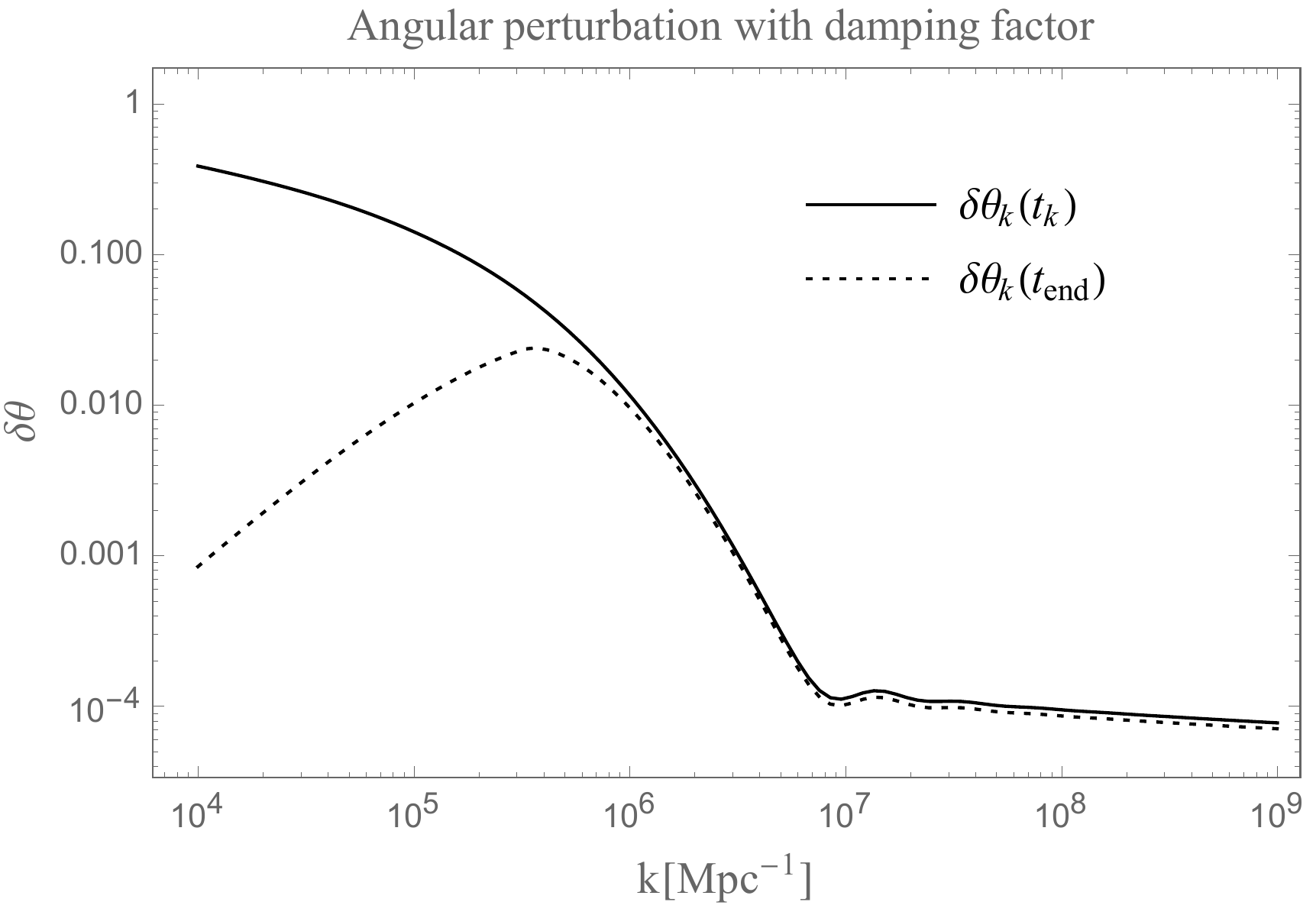}
 \caption{
Angular perturbation $\delta\theta_k$ for the LIGO-PBH using the delta-function window.
The solid line shows the angular perturbation at $t_k$. 
Because of the positive effective mass, the large-scale perturbation with $k<k_\tx{pbh}$ is damped (dotted line). 
}
 \label{thetadel}
\end{figure}

To explain how our model produces the sharp peak in the power spectrum, 
we plot the angular perturbation $\delta\theta_k$ at the horizon-crossing time $t_k$ and $t_\tx{end}$ as a function of $k$ in \Fig{thetadel} for the LIGO-PBHs with delta-function window.
Here $\delta\theta_k\fn{t_\tx{end}}$ is written as
\al{
\delta\theta_k\fn{t_\tx{end}}
&=\delta\theta_k\fn{t_{k}}	R_k
=
	\prn{H\fn{t_k}\ov 2\pi {\varphi_0}\fn{t_k}}
	R_k
	\propto\sqrt{\mc P_{\zeta}\fn{k,t_\tx{end}}}.
}
In \Fig{thetadel} it is seen that the shape of $\delta\theta_k(t_k)$ (solid line) mainly follows $\propto \ab{\Phi_0}^{-1}$ in \Fig{Zeromodedynamisc}. Using the damping factor $R_k\simeq \prn{k/ k_\tx{pbh}}^{3\ov2}$ in \Eq{Dampsim}, we get the damped angular perturbation (dotted line). 
Since the damping effect mainly works for $t<t_\tx{pbh}$, the small-scale perturbations with $k>k_\tx{pbh}$ do not suffer from the damping effect.
 
\subsection{Power spectrum of the curvature perturbations and PBH production}

In order to calculate the PBH mass spectrum, we should fix two more parameters, the curvaton energy ratio $r$ and the misalignment angle $\theta_i$, which affect the curvature spectrum as $\mathcal P_\zeta\propto \prn{2r/(4+3r)}^2\theta_i^{-2}$. 
We determine $r$ and $\theta_i$ in the following way.
For the given PBH density ${\Omega_\tx{PBH}/ \Omega_\tx{DM} }=1$ (DM-PBH) or ${\Omega_\tx{PBH}/ \Omega_\tx{DM} }=10^{-3}$ (LIGO-PBH), we search the allowed parameter region of $r$ and $\theta_i$ to achieve the required ${\Omega_\tx{PBH}/ \Omega_\tx{DM} }$.
We list the typical values of $r$ and $\theta_i$ for each window function in Table~\ref{rthetaTable}. 
The different window functions require the different values of $\mathcal P_\zeta$ to produce enough amount of PBHs. 
As in \Eq{windowChoie}, the Gaussian window function requires the largest $\mathcal P_\zeta$ to produce a sufficient number of PBHs and the top-hat window function requires the smallest.  

\begin{table}[t]
\begin{center}
\caption{Table for the typical value of $r$ and $\theta_i$}
   \label{rthetaTable}
\begin{tabular}{|l||c|c|} \hline
$(r,\theta_i)$ 		& LIGO-PBH	& DM-PBH \\ \hline \hline
Gaussian window  	& \gray{(1.0~,~0.0415)}	 	& \gray{(1~,~0.0380)}\\\hline
delta-function window	& (0.5~,~0.0524)		&(0.5~,~0.0481) \\\hline
Top-hat window  	&(0.5~,~0.0898)			& (0.5~,~0.0813)\\   \hline
\end{tabular}
  \end{center}
  
We list the required values of $(r, ~\theta_i)$ to produce enough amount of PBHs for each set of curvaton parameters in~\Eq{curvatonParames}. 
We estimated them for the three window functions including the effect of non-Gaussianity. 
\end{table}

\begin{figure}[ht]
 \centering 
  \includegraphics[width=.65\textwidth]{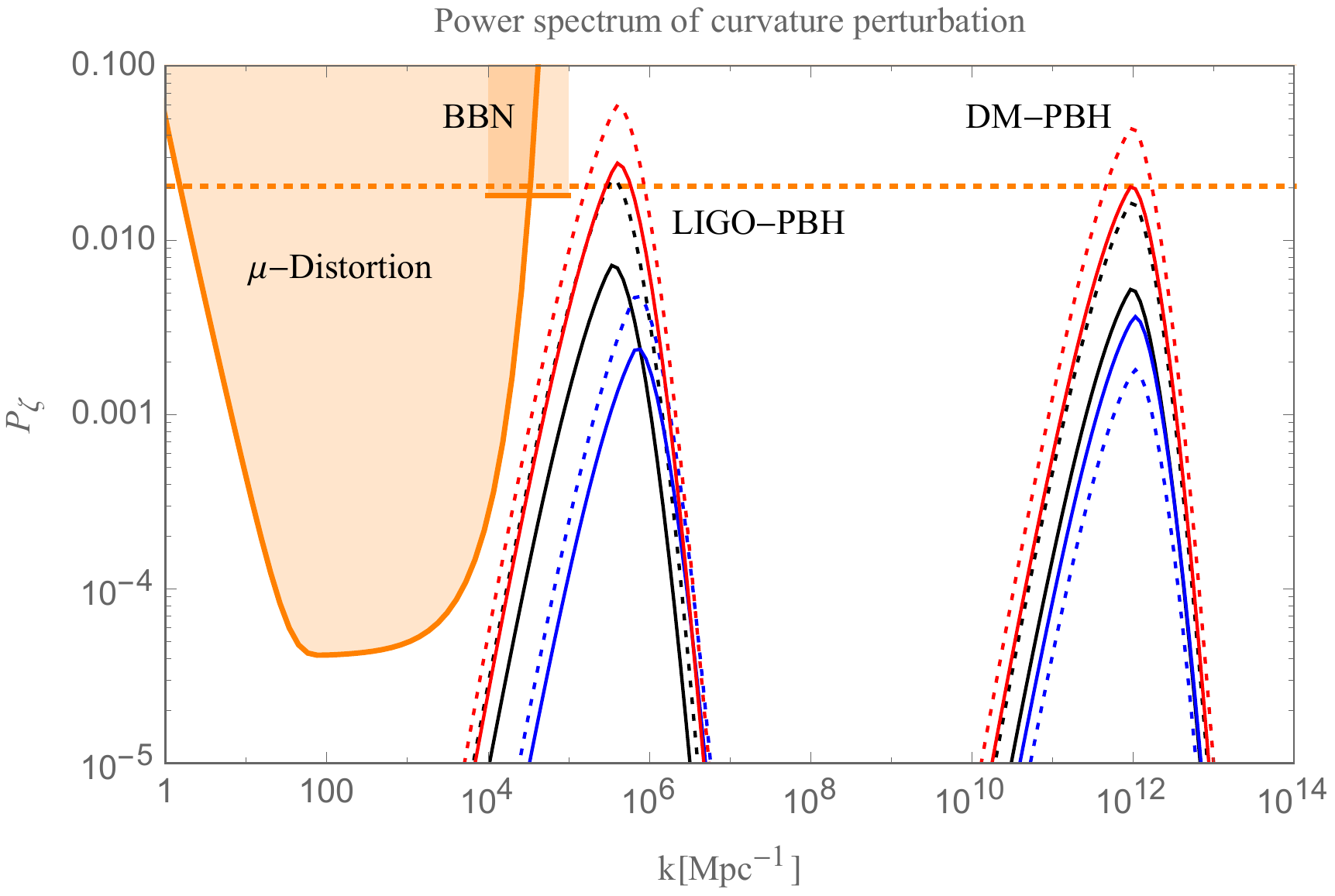}
 \caption{
 Power spectra $\mc P_\zeta$ for DM-PBH (peak around $k=10^6\,\Mpc^{-1}$) and LIGO-DM (peak around $k=10^{12}\,\Mpc^{-1}$).  
 We plot the power spectra for the delta-function window (black lines), top-hat window function (blue lines)  and Gaussian window function (red lines).   
 The solid line and dotted line show the spectra with/without non-Gaussianity. 
 The constraint from $\mu$-distortion\cite{0004-637X-758-2-76} by the COBE/FIRAS \cite{Fixsen:1996nj} and the BBN constraint~\cite{Inomata:2016uip} are also shown by orange shaded regions. 
The constraint from the curvaton fluctuations is shown by the orange dotted horizontal line [see, \Eq{isoConst}].
 }
 \label{powers}
\end{figure}

For fixed ${\varphi}_0 \fn{t}$, $r$ and $\theta_i$, we calculate the power spectrum of the curvature perturbations $\mc P_{\zeta}\fn{k}$ [\Eq{Pzetaformula}] which is shown in \Fig{powers}.
We plot $\mc P_{\zeta}(k)$ for the three window functions: Gaussian window (red lines), delta-function window (black lines) and top-hat window (blue lines).
We compare the results with/without the non-Gaussianity effect (solid lines/dotted lines).
Since we fix the PBH fraction, the non-Gaussianity effectively lowers the required $\mathcal P_\zeta$ as explained in \Eq{Deampnongau}.

The power spectrum of the curvature perturbations is constrained by CMB $\mu$-distortion~\cite{Fixsen:1996nj,0004-637X-758-2-76,Kohri:2014lza} and big bang nucleosynthesis(BBN)~\cite{Jeong:2014gna,Nakama:2014vla,Inomata:2016uip}.
The curvature perturbations on small scales dissipate through the Silk damping into the radiation. 
Such energy transfer distorts the spectrum of the CMB ($\mu$-distortion).
When the large curvature perturbations re-enter the horizon during BBN, they modifies the freeze-out value of the neutron-proton ratio. 
This modification is constrained by the observed ${}^4\tx{He}$ abundance~\cite{Aver:2015iza}.
In \Fig{powers} it is shown that our model avoids those constraints on $\mc P_{\zeta}\fn{k}$. 
This is because our model produces the steep spectrum of $\mc P_{\zeta}\fn{k}$ on the small scales by the positive effective mass as can be seen in \Eq{Dampsim}.

\begin{figure}[ht]
 \centering 
 \includegraphics[width=.65\textwidth]{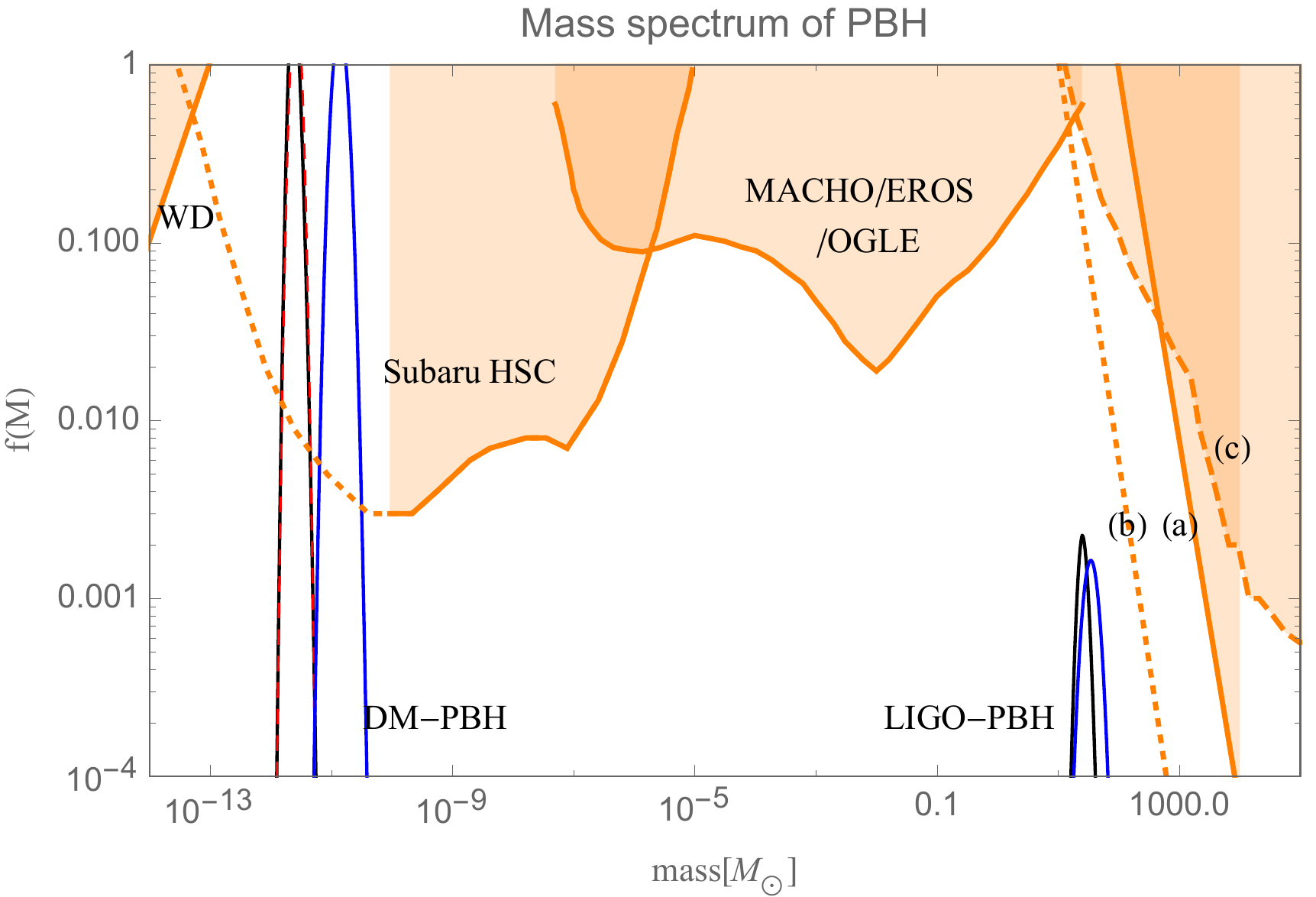}
 \caption{
 PBH mass spectra $f\!(\!M\!)$ of DM-PBH around $10^{-12}\msol$ and LIGO-DM around $30\msol$ for the delta-function window (black lines) and top-hat window function (blue lines).
 The region labeled ''WD'' shows the constraint from the white dwarfs~\cite{Graham:2015apa}.
The regions labeled ''Subaru/HSC''~\cite{Niikura:2017zjd} and ''MACHO/EROS/OGLE''~\cite{Allsman:2000kg,Tisserand:2006zx,Wyrzykowski:2011tr} are constraints from microlensing experiments (the dotted line shows the the constraint without ''wave effect''\cite{Inomata:2017vxo,Ohanian1974,Bliokh1975,Bontz1981}).
''(a)'' and  ''(b)'' are the constraints from the accretion effect on CMB depending on different assumptions~\cite{Ali-Haimoud:2016mbv}.
''(c)'' is the constraint by the x-ray emission from accretion gas around PBHs \cite{Inoue:2017csr}.
 We also give an example of the log-normal fitting  (red dashed line) given by
 $f(M)=1.5^2\exp(-19\ln^2(M/(2.6\times 10^{-12}\msol))~ )$
 .
 }
 \label{PBHsDistri}
\end{figure}

Next we discuss the the PBH mass spectrum $f\fn{M}$, 
{which is numerically calculated by \Eq{fMformula}}.
The PBH mass spectra in our model are shown in~\Fig{PBHsDistri} for the delta-function (black lines) and top-hat (blue lines) window functions.
Since we fix the PBH mass density, the mass spectrum depends only on the choice of the window function, which mainly changes the peak width and the peak mass of each distribution.

The PBH distribution is severely constrained by mainly two types of observations.
\ie{
\item 
	For PBHs with $10^{-10}-10\msol$, microlensing observations put the severe constraints.
A massive compact object works as a gravitational lens and amplifies the brightness of the background stars on the line of sight (microlensing).  
Subaru/HSC surveyed the Andromeda galaxy (M31) \cite{Niikura:2017zjd} and MACHO/EROS/OGLE surveyed the Large and Small Magellanic Cloud \cite{Allsman:2000kg,Tisserand:2006zx,Wyrzykowski:2011tr} and put the constraints on the abundance of the massive compact objects with monochromatic mass.
They have excluded the DM-PBH scenario in the wide mass range $(10^{-10}-10)\msol$. 
The Subaru/HSC observation~\cite{Niikura:2017zjd} obtained the constraint for smaller mass than $ 10^{-10}\msol$.
However, as the authors in \cite{Niikura:2017zjd} mentioned,  the constraint would become much weaker since the observational wavelength is comparable to the Schwarzschild radius of the lensing objects (wave effect)~\cite{Inomata:2017vxo,Ohanian1974,Bliokh1975,Bontz1981}.
Thus, we denotes the Subaru constraint for mass less than $10^{-10}M_\odot$ by the  dotted line in \Fig{PBHsDistri}. 	
\item For large mass PBH $>10^2\msol$, 
{some observations put constraints.
Severe constraints are given by the mass accretion process onto PBHs.}
In the early universe, the gas accretion injects energy into CMB, which affects the CMB anisotropies and leads to the constraint~\cite{Ali-Haimoud:2016mbv}.
Freely-floating black holes in the interstellar medium are constrained by the X-ray emission from its accretion gas \cite{Inoue:2017csr}.
{
There are other constraints from a stability of star cluster near the center of the ultra-faint dwarf galaxy \cite{Brandt:2016aco} and 
FRB lensing \cite{Munoz:2016tmg}.
}
\item Other constraints: For PBHs with smaller mass than $10^{-13}\msol$, the observation of white dwarfs gives a constraint. 
	When PBHs collide with white dwarfs, they heat them by the dynamical friction and cause explosion~\cite{Graham:2015apa}.
}

Although the PBH mass spectrum $f\fn{M}$ has been already severely constrained, our model avoids these constraints by the sharp peak spectrum. 
For extended mass functions like our case, it is nontrivial to compare them with the observational constraints which are derived assuming monochromatic mass functions.  
We have confirmed that the PBH mass spectrum in \Fig{PBHsDistri} avoids constraints using the treatment in \cite{Carr:2017jsz,Inomata:2017okj}.  


\subsection{Secondary gravitational wave}

We estimate the density of the secondary gravitational waves $\Omega_{GW}$ by numerically integrating Eq.~(\ref{eq:gravitational_wave}).
In estimating $\Omega_{GW}$ we assume that the curvaton has already decayed when the peak wavelength of the density perturbation spectrum re-enters the horizon. 
For the typical parameters, the curvaton decays at $T_\sigma\simeq 10^7\,\GeV$ and DM-PBHs are produced at $10^5\,\GeV$ and LIGO-PBHs at $50\,\MeV$.\footnote{
The curvature perturbations due to the curvaton before the decay requires a different treatment for calculating the $\Omega_{GW}$. 
Including 
the contribution before the curvaton decay, $\Omega_{GW}$ of the DM-PBH, could slightly change the spectrum shape but the typical value of $\Omega_{GW}$ would not change.
}

\begin{figure}[t]
 \centering 
 \includegraphics[width=.65\textwidth]{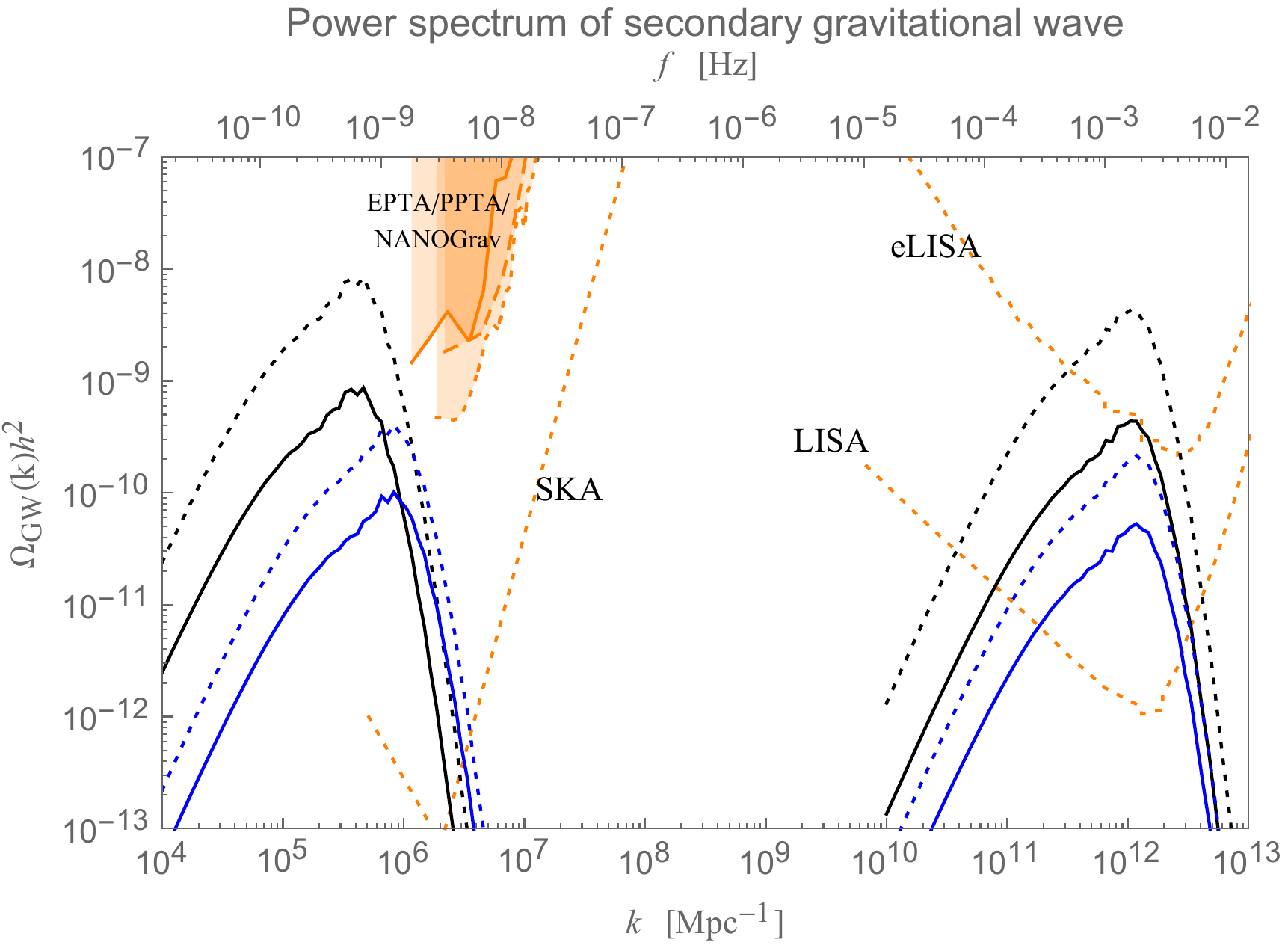}
 \caption{
 The differential energy density of secondary gravitational waves. 
 The power spectra of LIGO-PBHs (peak around $10^5\,\Mpc^{-1}$) and DM-PBHs (peak around $10^{12}\,\Mpc^{-1}$) are plotted for the delta-function  (black) and top-hat  (blue) window functions.
 The solid and dotted lines are the power spectra with/without non-Gaussianity.
 The constraint from the current pulsar timing array experiments is shown by the orange shaded region:EPTA \cite{Lentati:2015qwp} (solid line), NANOGrav \cite{Arzoumanian:2015liz} (dashed line) and PPTA \cite{Shannon:2015ect} (dotted line).
 The future SKA experiment~\cite{Moore:2014lga} has much improved sensitivity shown by red dotted line around $10^{6}\,\Mpc^{-1}$.
The future constraints by the eLISA/LISA are also shown by the red dotted lines around $10^{12}\,\Mpc^{-1}$.
}
 \label{2ndGWfig}
\end{figure}

Pulsar Timing Array (PTA) experiments put severe constraints on stochastic GWs \cite{Lentati:2015qwp,Arzoumanian:2015liz,Shannon:2015ect} with a wave number of around $10^6\,\Mpc^{-1}$ which is also shown in~\Fig{2ndGWfig}.
Our model for LIGO-PBHs avoids the current PTA constraints because the spectrum has a sharp peak. 
The future experiments can detect the GWs predicted by the curvaton model.
For example, the Square Kilometer Array (SKA) would improve the sensitivity of stochastic GWs around $10^6\,\Mpc^{-1}$~ \cite{Janssen:2014dka,Moore:2014lga}.
For the stochastic GWs around $10^{12}\,\Mpc^{-1}$, eLISA/LISA would put the constraints for DM-PBH scenario~\cite{Moore:2014lga}.

\subsection{Constraint from curvaton perturbations}
\label{rthetaConst}

Although the smaller $\theta_i$ enhances $\mathcal P_\zeta\propto {\theta_i}^{-2}$, $\theta_i$ is constrained from below by the condition that the energy perturbation should not be larger than the mean value, i.e.,  ${\delta\rho_\sigma/ \rho_\sigma} < 1$.
From \Eq{densityperturb} the condition can be rewritten as
\al{
1&>	\prn{\delta\rho_\sigma \ov \rho_\sigma}^2
	=\prn{4+3r\ov r}^2\mathcal P_\zeta\fn{k}.
	\label{Isocurvcondition}
}
In this paper, we assume that the PBH production during the radiation-dominated era and $r<1$, which leads to
\al{
\mathcal P_\zeta\fn{k}&< {1\ov 7^2}\simeq 0.0204.
\label{isoConst}
}
The condition \Eq{Isocurvcondition} does not hold in the Gaussian window function case, which requires the large $\mathcal P_\zeta$ to produce enough PBHs. 
In \Fig{powers}, we can see that $\mathcal P_\zeta$ obtained by adopting the delta-function and top-hat window functions avoids the constraint~\Eq{Isocurvcondition} with/without non-Gaussianity.
		
\section{Conclusion} \label{sec_conclu}

In this paper,  we have studied the PBH formation in the axion-like curvaton model~\cite{Ando:2017veq}.
We have modified the original model by introducing a coupling with an inflaton field, which leads to a sharp peak in the power spectrum of the curvature perturbations.
The coupling with inflaton also enables us to choose the mass scale of produced PBHs without the tuning of an initial field value or the curvaton decay rate.
We have also evaluated the non-Gaussianity effect of the curvaton and uncertainty from the choice of the window functions. 
It has been found that our model produces enough PBHs as LIGO events ($30\msol$) and DM ($10^{-12}\msol$).
	
Furthermore, our calculation has shown that our model is consistent with the current constraints on the curvature perturbation $\mc P_\zeta\fn{k}$ (see \Fig{powers}),  on the PBH mass function $f\fn{M}$ (see \Fig{PBHsDistri}), and on the secondary GW $\Omega_{GW}$ (see \Fig{2ndGWfig}).
Next-generation observations would verify our model through secondarily produced GWs by PTA experiments like  SKA and/or by space gravitational wave interferometers like LISA and eLISA. 

\section*{Acknowledgements}
\small\noindent
This work was supported by JSPS KAKENHI Grants No. 17H01131 (M.K.) and No. 17K05434 (M.K.), MEXT KAKENHI Grant No. 15H05889 (M.K.), World Premier International Research Center Initiative (WPI Initiative), MEXT, Japan (K.A., M.K., H.N.), JSPS Research Fellowships for Young Scientists Grant No. 18J21906 (K.A.), and the Advanced Leading Graduate Course for Photon Science (K.A., H.N.).

\appendix

\section{Equation of motion of the curvaton}
\label{pertcalc}

During inflation, the potential of the present model is given by
\al{
V_\Phi=
	{\lambda\ov 4}\prn{\ab{\Phi}^2-{{v^2\ov 2 } }}^2
	+g\phi^2\ab{\Phi}^2
	-v^3 \epsilon \prn{\Phi+\Phi^*}.
}
We decompose the complex scalar $ \Phi$ into a homogeneous part $\Phi_0$ and its perturbation $ \Phi_1$.
Using $\Phi_0$ and $ \Phi_1$ the action on the FLRW space-time is written as
\al{
S&=\int \df^4 x\sqrt{-g}\prn{-\prn{\Phi_0+ \Phi_1} _{,\mu}\prn{\Phi_0+ \Phi_1} _{,\nu}^* g^{\mu\nu}-V_\Phi\sqbr{{\Phi_0+ \Phi_1}}}
\nn&=
	\int \df \eta \df^3 x a^4\prn{{1\ov a^2}(\ab{\Phi_0+ \Phi_1}^{\pr 2}-\ab{\nabla \Phi_1}^2 )-V_\Phi \sqbr{{\Phi_0+ \Phi_1}}}
}
with $\sqrt{-g}= a^3$ and  $\Phi'={\p\Phi/ \p\eta}=a{\p\Phi/\p t}=a\dot\Phi$. The equation of motion is 
\al{
0&={\p \ov\p \eta }{\delta S\ov \delta  \Phi^{\pr *}}
	+{\p \ov\p x^i }{\delta S\ov \delta \Phi_{,i}^*}
	-{\delta S\ov \delta \Phi^*}
\nn&=
	{\p a^2 \prn{\Phi_0+\Phi_1}'\ov\p \eta }
	{-} {\p a^2 \Phi_{1,i}\ov\p x^i}
	+a^4{\p V_\Phi \sqbr{{\Phi_0+\Phi_1}}\ov \p \Phi^*}
\nn&=
	a^2\sqbr{
	\prn{\Phi_0+\Phi_1}''
	+2aH\prn{\Phi_0+\Phi_1}' 
	{-}\nabla^2\Phi_1
	+a^2{\p V_\Phi \sqbr{{\Phi_0+\Phi_1}}\ov \p \Phi^*}
	},
	\label{EoMc}
}
where
\al{
{\p V_\Phi\ov \p  \Phi^*}
	=
	\prn{g\phi^2
    	-{\lambda\ov 2}	\prn{{{v^2\ov 2 }}-\ab{\Phi}^2}
        }\Phi
	-\epsilon v^3  .
    \label{delpote}
}
From \Eq{EoMc} the homogeneous part satisfies 
\al{
0&=\Phi_0''
	+2aH\Phi_0' 
	+a^2\sqbr{
	\prn{g\phi^2-{\lambda\ov 2}\prn{{{v^2\ov 2 }}-\ab{\Phi_0}^2}}\Phi_0
	-\epsilon v^3 
	}.
	\label{Csol}
}
Note that $\Phi_0$ rolls down the Higgs-like potential in the direction of $\arg\fn{\Phi_0}=0$ with $\epsilon>0$.

Using \Eq{Csol} in \Eq{EoMc}, the perturbation $\Phi_1$ satisfies 
\al{
0&= \Phi_1''
	+2aH\Phi_1'
	{-}\nabla^2\Phi_1
	+a^2\prn{ {\p V_\Phi\sqbr{\Phi_0+\Phi_1}\ov\p \Phi^* } -{\p V_\Phi\sqbr{\Phi_0}\ov\p \Phi^* } }
\nn&
	=a^2\sqbr{
	\ddot\Phi_1
	+3H\dot\Phi_1
	{-} a^{-2}\nabla^2 {\Phi_1}
	+\prn{
    u{\Phi_1}^*
	+s{\Phi_1}
	+w[\Phi_0,\Phi_1]
		}
	}
\nn
u&={\lambda \ov 2}\Phi_0^2 
\qcq
s=g\phi^2
	-{\lambda v^2\ov 4}
    +{\lambda}\ab{\Phi_0}^2
\qcq
w=\frac{\lambda}{2}\prn{\Phi_1\ab{\Phi_1}^2+2\Phi_0\ab{\Phi_1}^2 +\Phi_0^*\Phi_1^2}.
}
Hereafter we only consider the terms linear in $\Phi_1$.
With the Fourier transformation and the diagonalization, the equation of motion is given by
\al{	
&
	\sqbr{
    \p^2_t +3H\p_t +a^{-2}k^2
    +\pmat{s &u\\ u^*&s}
    }
	\pmat{\Phi_{1;k}\\ \Phi_{1;k}^*}=0,\nn
&
	 \sqbr{\p^2_t +3H\p_t +a^{-2}k^2+
     \pmat{s+u \\ s-u}
     }
     {1\ov \sqrt{2}}
	 \pmat{\Phi_{1;k}+\Phi_{1;k}^* \\- \Phi_{1;k}+ \Phi_{1;k}^*}=0.
     \label{EoMpert}
}
where we use that $u$ is real in our calculation. 
Rewriting $\Phi$ as
\al{
\Phi
	={1\ov\sqrt{2}}(\varphi_0+\varphi)e^{i\sigma/\varphi_0}
    = {1\ov \sqrt{2}}\varphi_0
    +\prn{
    	{\varphi\ov \sqrt{2}} +{i\sigma \ov \sqrt{2}}
        }
    =\Phi_0+\Phi_1,
}
we get effective mass of $\varphi$ and $\sigma$ as
\al{
m_\varphi^2
&=s+u
  =g\phi^2
  	-{\lambda v^2\ov 4}
    +\frac{3}{2}{\lambda}\ab{\Phi_0}^2
  \qcn
 \tilde m_\sigma^2
&=s-u
  =g\phi^2
  	-{\lambda v^2 \ov 4}
    +\frac{1}{2}{\lambda}\ab{\Phi_0}^2.
}

When the perturbation is super horizon $aH\gg k$, assuming that Hubble parameter is constant, \Eq{EoMpert} is written as  
\begin{equation}
   [\partial_t^2 +3H +m_\text{eff}^2]X =0,
\end{equation}
where $X=\varphi, ~\sigma$ and $m^2_\text{eff} = m^2_\varphi ~(\tilde m^2_\sigma)$ for $X=\varphi ~(\sigma)$.
The solution can be written as $X \propto e^{i\omega_k t}$ with $\omega_k$ given by
\al{
&
	\omega_k ={3\ov 2}i H
    \prn{
    	1\pm \sqrt{1-\prn{\frac{2m_\tx{eff}}{3H} }^2}
        }
	.
}

In $m_\tx{eff}>0$, the perturbations are strongly damped as
\al{
{\p \ln\ab{\Phi_{1;k}}\ov\p t}=-{3\ov 2} H\re 
	\sqbr{1-\sqrt{1-\prn{\frac{2m_\tx{eff}}{3H} }^2}}.
\label{damphigh}
}

\section{The curvaton energy ratio}
\label{curvatondynamics}
 
 In this section, we summarize the curvaton dynamics in the radiation-dominated era \cite{Kawasaki:2012wr}. 
 We evaluate the energy ratio $r={\rho_\sigma/ \rho_I}$ when both curvaton and inflaton decay into the radiation. 
We define several characteristic times: $t_\text{end}$ when inflation ends, $t_{R}$ when the inflaton decays, $t_{\sigma;\tx{osc}}(H=m_\sigma)$ when the curvaton starts to oscillates and  $t_{\sigma;\tx{dec}}$ when the curvaton decays. 
We focus on the case $t_\tx{end}<t_{\sigma:\tx{osc}}<t_\tx{R}<t_{\sigma;\tx{dec}}$  to achieve the large $r$ value. 
We can rewrite this condition as $m_\phi>m_\sigma>\Gamma_\tx{R}>\Gamma_\sigma$ where $\Gamma_\tx{R}$ and $\Gamma_\sigma$ are decay rates of the inflaton and the curvaton, respectively.
Then, the energy ratio is given by~\cite{Kawasaki:2012wr}
\begin{equation}
   r= \frac{T_\text{R}}{T_\sigma}\,\frac{v^2\theta_i^2}{6M_{pl}^2},
\end{equation}
where $T_\text{R}$ and $T_\sigma$ are the reheating temperature after inflation and the temperature at $t_{\sigma;\tx{dec}}$.
Assuming the instantaneous decay, 
\al{
\Gamma_\tx{R}&=\sqrt{{\pi^2\ov 30}(\g_*T^4_\tx{R}/3\mpl^2)}
\qcq
\Gamma_\sigma=\sqrt{{\pi^2\ov 30}(\g_*T^4_\sigma/3\mpl^2)}.
}
$T_\sigma$ should be larger than several MeV because otherwise the curvaton decay spoils the success of BBN. 
We also require that the curvaton decays before the PBH formation, $T_\sigma>10^5\,\GeV$(PBH-DM) and $T_\sigma>30\,\MeV$(LIGO-PBH) in \Eq{PBHmasses}. 
Note that in our typical parametrization, 
${v^2\theta^2/ (6\mpl^2 )}\sim 10^{-6}$ 
for both PBH-DM and LIGO-PBH cases. 
Thus, $r=\mc O\fn{0.1}$ in Table~\ref{rthetaTable} requires the $\prn{T_\tx{R}\ov T_\sigma}\sim 10^5$.  
For example, with $T_\tx{R}=10^{12}\,\GeV$, the curvaton decay occurs at $T_\sigma\sim 10^7\,\GeV$.
Both the DM-PBH and the LIGO-PBH scenarios are available, but only for the limited parameter regions.

With typical curvaton decay parametrization $\Gamma_\sigma=\prn{\kappa^2/ 16\pi}\prn{m_\sigma^3/ {v}^2}$ with $\kappa$ being a constant, we can check the validity of our assumption $m_\phi>m_\sigma>\Gamma_\tx{R}>\Gamma_\sigma$.
Taking $T_\sigma\sim 10^{7}\,\GeV$, we get 
$\Gamma_\sigma\sim T_\sigma^2/\mpl\sim 10^{-2}\,\GeV$, 
$m_\sigma\sim \prn{16\pi v^2\Gamma_\sigma/ \kappa^2}^{1/3}\sim10^{10}\,\GeV$.
With $T_\tx{R}=10^{12}\,\GeV$ and $\Gamma_\tx{R}\sim T_\tx{R}^2/\mpl\sim 10^6\,\GeV$, we get
\al{
m_\phi(\sim 10^{13}\,\GeV)>
m_\sigma(\sim 10^{10}\,\GeV)>
\Gamma_\tx{R}(\sim 10^{6}\,\GeV)>
\Gamma_\sigma(\sim10^{-2}\,\GeV).
}

\small
\bibliographystyle{apsrev4-1}
\bibliography{Ref.bib}

\end{document}